\documentclass[pra,showpacs,superscriptaddress,preprintnumbers,twocolumn,amsmath,amssymb,floatfix]{revtex4}

\usepackage{graphicx}
\usepackage{dcolumn}
\usepackage{bm}

\begin{document}

\title{{\rm\small\hfill (Phys. Rev. A, accepted)}\\
        Image states in metal clusters}

\author{Patrick Rinke}
\altaffiliation{Present address: Fritz-Haber-Institut der
Max-Planck-Gesellschaft, Faradayweg 4--6, 14195 Berlin, Germany}
\affiliation{Department of Physics, University of York, Heslington, York YO10
             5DD, United Kingdom }
\author{Kris Delaney}
\altaffiliation{Present address: Department of Physics, University of Illinois
                at Urbana-Champaign, Urbana IL 61801-3080, USA}
\affiliation{Department of Physics, University of York, Heslington, York YO10
             5DD, United Kingdom }
\author{P. Garc\'{\i}a-Gonz\'alez}
\altaffiliation{Present address: Departamento de F\'{\i}sica Fundamental, 
                UNED, Apartdado 60141, E-28080 Madrid, Spain}
\affiliation{Departamento de F\'{\i}sica de la Materia Condensada, Universidad
             Aut\'{o}noma de Madrid, E-28049, Madrid, Spain}
\author{R. W. Godby}
\affiliation{Department of Physics, University of York, Heslington, York YO10
             5DD, United Kingdom }

\date{\today}

\begin{abstract}
The existence of image states in small clusters is shown,
using a quantum-mechanical many-body approach. We present image state energies and
wave functions for spherical jellium clusters up to 186 atoms, calculated in 
the $GW$ approximation, where $G$ is the Green's function, and $W$ 
the dynamically screened Coulomb
interaction, which by construction contains the 
dynamic long-range correlation effects that give rise to image effects. 
In addition we find that image states are also subject to quantum confinement.
To extrapolate our investigations to clusters in the mesoscopic size range, we
propose a semiclassical model potential, which we test against our full $GW$
results.

\end{abstract}


\maketitle

%
%
\section{Introduction} 
Image states are highly extended, excited electronic states that occur 
predominantly 
at the surface of a polarizable material when an extra electron is added to the
system. Electrons in such an image state 
feel the attractive force of the charge induced in the material
even far away from the surface due to the extremely long-ranged correlation of
the Coulomb potential.

In the past, research on image states was mostly devoted to metal surfaces, 
both experimentally  \cite{exp_1} 
and theoretically \cite{Silkin/Chulkov:1999_s}. 
Recently, however, studies have also been extended to
nanotubes \cite{Granger/Kral/Sadeghpour/Shapiro:2002} and
metallic nanowires on surfaces \cite{Hill/McLean:1999}.
Unlike surfaces, isolated 
nanoclusters are not stationary and image states can therefore only be resolved
indirectly with the experimental techniques currently available.
By measuring the capture cross section of low-energy electrons, for instance,
Kasperovich {\it et al}. 
were able to identify a clear signature of image effects in free sodium
clusters of 4 nm radius \cite{Kasperovich/Kresin:2000}.

In the context of water clusters, polar molecules, and clusters of rare gas 
atoms, excess electron states have
 been widely discussed in the literature
\cite{Stampfli:1995}. 
The electron-electron interaction in these clusters is typically
included using quasiclassical dielectric screening, 
which becomes justified in the mesoscopic regime but is not parameter-free 
\cite{dcontinuum_1,dcontinuum_2,Mo/Sung/Ritchie:1991}.
For smaller clusters, the interaction of the excess electron with the cluster 
has been modeled using electron-atom pseudopotentials, with the ground state 
geometry from  molecular dynamics \cite{MD}; image states were not included in
the study.

While the effect of the cluster polarization potential on the 
scattering \cite{scattering_cs}
and capture \cite{capturing_cs}
cross section has been studied with a variety of different approaches,
 we will
focus in this paper specifically on the bound states that arise from 
the interaction of the excess electron with its image charges.
We report first-principles calculations for jellium clusters with
sodium densities as a prototype system for isolated nanoclusters.  The
Coulomb interaction between all valence electrons is taken into account in the  
framework of many-body perturbation theory.
The image state energies
and wave functions were obtained using the 
$GW$ approximation \cite{Hedin:1965}, which has proven to be very successful for the 
description of image effects 
\cite{impot,Silkin/Chulkov:1999_s, Fratesi/Brivio/Rinke/Godby} 
and other quasiparticle properties 
\cite{Aulbur/Joensson/Wilkins:1999}.
Our calculations predict 
the existence of image states in these zero-dimensional nanostructures
even down to relatively small cluster sizes.
%
%

\begin{figure*}
\begin{minipage}[b][5cm][c]{0.64\linewidth}
  \includegraphics[height=5cm]{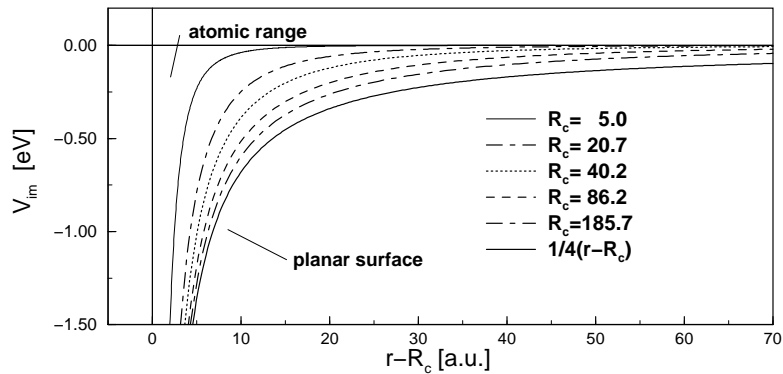}
\end{minipage}
\begin{minipage}[b][5cm][c]{0.34\linewidth}
  \caption{\label{im:sphimpot} 
              The classical image potential, $v_{\rm im}^c(r)$ [Eq. \ref{im:pot}], of a
	      solid sphere with $\varepsilon = 1000$ becomes more accommodating with
	      increasing radius $R_c$ (in a.u.) and is bound by the  
	      planar surface (solid line to the right) and the atomic
	      limit (solid line to the left).}
\end{minipage}
\end{figure*}

By way of illustration, the classical image potential outside a neutral 
solid sphere with radius $R_c$ and 
dielectric constant $\varepsilon$ has the form \cite{footnote:au}
\begin{equation}
\label{im:pot}
  v_{\rm im}^c(r)=\sum_{n=1}^\infty 
            \frac{-n(\varepsilon -1)}{2n(\varepsilon+1)+2}
            \frac{R_c^{2n+1}}{r^{2n+2}} 
	    \approx
	    \frac{-R_c^3}{2r^2(r^2-R_c^2)}
\end{equation}
for $r>R_c$. The last expression illustrates the limit of a 
perfectly conducting sphere  ($\varepsilon \rightarrow \infty$);
the familar image potential of a flat surface, $-1/4z$ where
$z=r-R_c$, is recovered for large cluster radii.

The image potential for a solid sphere (\ref{im:pot}) decays asymptotically
as $-1/r^4$ and thus much more
rapidly than the image potential of the planar surface. 
However, in a region of size of order $R_c$ just outside the surface, the much
more accommodating flat-surface form prevails (Fig. \ref{im:sphimpot}).

As previously reported \cite{Granger/Kral/Sadeghpour/Shapiro:2002}
the image potential of a metallic tube with radius $R_t$ decays
asymptotically as $ - 1/[r R_t \ln(r/R_t)]$
and is thus effectively situated closer to 
the flat-surface limit in Fig. \ref{im:sphimpot}. For both the cluster and the
tube, the image potential
depends on the radius of the nanostructure. For clusters, this
dependence is considerable, as Fig. \ref{im:sphimpot} illustrates, and
consequently has a strong effect on the binding energy and the wave functions of
the image states, as we will show in the following.

%
%
\section{Computational Approach} 

The quasiparticle energies and wave functions are formally the solution of the
quasiparticle equation
\begin{equation}
\label{qpe:generic}
  \hat{H}_0\:\psi_n({\bf r}) +
  \int d{\bf r'} \Sigma({\bf r},{\bf r'};\epsilon_n) \psi_n({\bf r'}) = 
  \epsilon_n \psi_n({\bf r})
\end{equation}
with the effective one-particle Hamiltonian,
$\hat{H}_0$, including the Hartree and the external 
potential. The non-local, dynamical self-energy $\Sigma({\bf
r},{\bf r'};\omega)$ contains the electron exchange and correlation effects
beyond the Hartree mean-field and is, in the $GW$ approximation, given
by $\Sigma=GW$~\cite{Hedin:1965}, where $G$ is the Green's function, and $W$ 
the dynamically screened Coulomb
interaction.

In the jellium clusters studied in this paper, the atomic nuclei are 
replaced by a homogeneous, 
positive background charge, $\rho^+(r)=\rho_0 \Theta(R_c-r)$, 
with $\rho_0=3/4\pi r_s$. The Wigner-Seitz radius $r_s$ is indicative of the
electron density of the material and we chose a value of $r_s$=4.0 for the 
jellium clusters with sodium densities presented here. 

The electrostatic potential created by the background
charge density, $\rho^+(r)$, is spherically symmetric and,
therefore, all cross section planes through the origin of the cluster
are equivalent. It
is thus sufficient to describe the system by two radial coordinates $r$
and $r'$ and one angular coordinate $\theta$ that denotes the angle between
the vectors ${\bf r}$ and ${\bf r'}$. The self-energy then assumes the much
simpler form
$  \Sigma(r,r',\theta;\omega)=\sum_{l=0}^{\infty} \left[ \Sigma_l^x(r,r')+
			     \Sigma_l^c(r,r';\omega)\right] P_l(\cos\theta)
$.  

The Legendre expansion coefficients of the exchange, $\Sigma_l^x$, and the
correlation part, $\Sigma_l^c$, of the self-energy are calculated directly,
thereby surpassing the need for an explicit treatment of the angular dependence.
We use a real-space and imaginary time representation 
\cite{spacetime}
to calculate the self-energy from the Kohn-Sham Green's function of a preceding
density-functional calculation in the local density approximation (LDA). The
expression for the self-energy on the real frequency axis is obtained by means
of analytic continuation \cite{spacetime}.

To obtain the quasiparticle energies and wave functions, the quasiparticle 
equation (\ref{qpe:generic}) is fully diagonalized in the basis of the LDA  
wave functions.  
The ionization potential and the electron affinities calculated in this way
agree well with available data from photoionization experiments 
\cite{Chandezon/Bjoernholm:1997} and 
are also in excellent agreement with the only previous $GW$ study on spherical
jellium clusters by Saito {\it et al} \cite{Saito/Louie}, in which a plasmon
pole model was used. 
The energy range in which 
surface and image states occur in jellium clusters, however, was not included
in their investigation. 
%
\section{Image States in Metal Clusters}

\begin{figure}
\includegraphics[width=1.0\columnwidth]{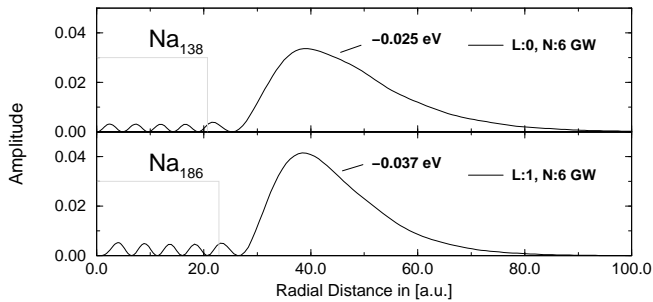}
\caption{\label{im:imstates_large}For each of the clusters Na${}_{138}$ and 
           Na${}_{186}$, a loosely bound 
           image state is found that predominantly resides in the vacuum
	   region outside the cluster and has very little overlap with the
	   cluster region. In the LDA, neither an image state 
	   nor a bound state with the same quantum numbers 
	   can be obtained. (The energies are referenced with respect to the
	   vacuum and the gray box marks the extent
	   of the clusters.)}
\end{figure}

\begin{figure}
\includegraphics[width=1.0\columnwidth]{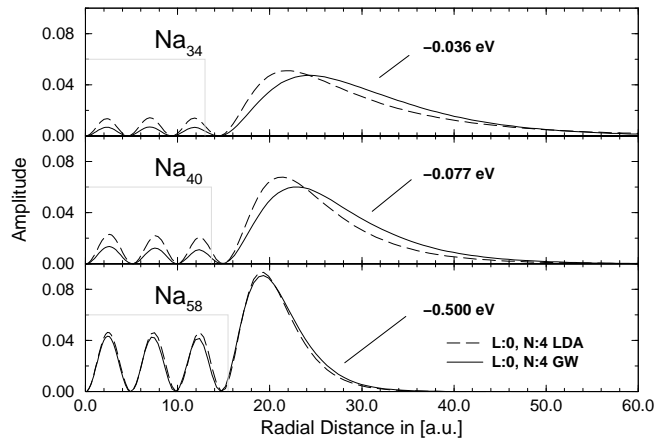}
\caption{\label{im:imstates_small} The highest image state (solid line) in
	    Na${}_{34}$ becomes more tightly
	    bound and more strongly localized with increasing cluster size,
	    Na${}_{40}$, and eventually evolves into a surface resonance in
	    Na${}_{58}$. For comparison, the LDA
	    wave functions have been included (dashed lines).}
\end{figure}

In Fig. \ref{im:imstates_large}
we present the highest image state
\cite{footnote:def_of_imstat} calculated for the clusters  Na${}_{138}$ and
Na${}_{186}$, respectively \cite{footnote:notation}. Both image states are 
very similar in character and 
extend extremely far into the vacuum having an almost insignificant 
overlap with the cluster. Most strikingly, however, is that the
corresponding LDA states bear no resemblance to the image state wave functions.
Due to the absence of long-range correlation 
effects in density-functional theory (DFT) the corresponding state in the LDA 
calculation becomes an unbound state that is scattered by the effective
potential.
This observation proves that a full diagonalization of the quasiparticle
Hamiltonian (\ref{qpe:generic}) is necessary,
because the LDA wave functions no longer provide a good description of the 
quasiparticle wave functions, as is the case for bulk 
\cite{GWdiag} and low lying
cluster \cite{Saito/Louie,
Pulci/Reining/Onida/DelSole/Bechstedt:2001} states.
The exchange-correlation potential in the LDA decays exponentially in the 
vacuum region as opposed to the $-1/r^4$ behavior of the image potential felt
by an extra electron. The similarity between these two potentials for small
clusters is coincidental and leads in certain cases to bound image states even
in the LDA (see Fig. \ref{im:imstates_small}).  

Owing to the much
more rapid decay of the image potential for a solid sphere (\ref{im:pot}), 
the energy band in which image states are found is reduced to $\sim$0.2 eV 
below the vacuum level for a small cluster compared 
to an energy range of approximately 1 eV found at metal surfaces 
\cite{Silkin/Chulkov:1999_s}.
The image state binding energies of $-$0.025 eV for Na${}_{138}$ and
$-$0.037 eV for Na${}_{186}$ are thus small in relation to 
energies of a few tenths of an
eV observed for sodium surfaces \cite{Silkin/Chulkov:1999_s}. 
For larger clusters, the overlap of the higher image 
states with the cluster region becomes negligible 
(see Fig. \ref{im:imstates_large}).
We therefore expect these states to be insensitive to any atomic structure of
the cluster, and to be long lived.

%

Focusing on the highest image state in the sodium cluster series, 
we now demonstrate that image states are subject
to quantum confinement effects. 
In Fig.~\ref{im:imstates_small}, we have illustrated the evolution of the
state \mbox{($L$=0,$N$=4)} with increasing cluster size.
In the smallest cluster, Na${}_{34}$, the image state is most narrowly bound 
at only $-$0.036 eV, whereas in the next larger cluster, Na${}_{40}$, it is 
localized closer to the surface and exhibits more overlap 
with the cluster itself. The same state has evolved into a surface resonance for the 
Na${}_{58}$ cluster and will eventually become an ordinary bound cluster state for
larger quantum dots. This size dependence of the image states 
results from a delicate interplay between the confinement of the cluster 
potential and the long range of the image potential.
If there was no overlap of the image state wave function with the cluster region,
as assumed in many classical approaches, then the image states in the size range
depicted in Fig.~\ref{im:imstates_small} would be identical, since the 
variation of the image potential itself can be regarded as negligible in this 
range (see Fig. \ref{im:sphimpot}).
In reality, however, the overlap with the cluster increases with cluster size, 
as can be seen in Fig.~\ref{im:imstates_small}, which is accompanied by an
increase in binding energy.
Subsequently, the electron in the image state 
will sample less and less of the long-ranging image tail. A reduced confinement by the external cluster
potential will therefore lead to a stronger confinement by the image potential
and thus to a stronger localization overall.

Furthermore, Na${}_{34}$ is the smallest cluster in the sodium series for which 
at least one image state is bound. 
This observation therefore positively answers the question that 
arises from Eq.~(\ref{im:pot}) and
Fig.~\ref{im:sphimpot} if there exists a minimum
cluster size for image effects to become important.

Experimentally it has recently been demonstrated 
by Kasperovich {\it et al}.  \cite{Kasperovich/Kresin:2000} that the size of 
sodium nanoclusters can be determined by 
measuring the contribution to the capture cross section that
arises from image effects.
In this context we therefore like to emphasize that, in contrast to surfaces, 
spherical nanostructures provide an extra parameter to 
tailor the binding energy and the shape of an images state. 

%
\section{Towards the Mesoscopic Size Range}
In the final part of this paper we will pursue the notion of incorporating
image effects into a suitable model potential in the framework of our DFT
calculations, in order to predict image states 
for clusters in the mesoscopic size range. Our full $GW$ results will thereby 
serve as a reference to establish the validity 
of the model and to give an estimate of its transferability. 

\begin{figure}
\includegraphics[width=1.0\columnwidth]{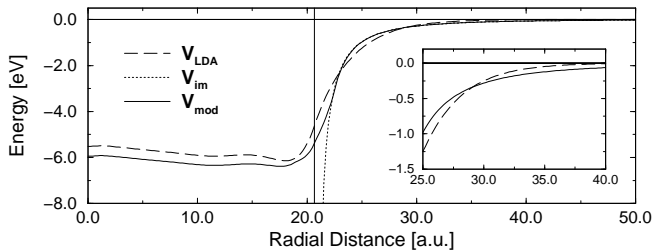}
\caption{\label{im:Na138impot} 
         The model potential of Eq. (\ref{im:modpot}) of a Na${}_{138}$ cluster
	 with $\varepsilon=1000$ (solid line) crosses over with the exponentially
	 decaying LDA-$v_{xc}$ potential (dashed line) and follows the inverse
	 power law decay of the classical image potential (Eq.
	 (\ref{im:pot})) (dotted line) by construction.
}
\end{figure}

Because the exchange-correlation potential in the LDA $v_{xc}(r)$
is proportional to a power
of the electron density, it will decay exponentially outside the cluster. The
image potential (\ref{im:pot}) on the other hand is long ranging since it
asymptotically follows an inverse power-law behavior 
that varies between $-1/r^4$ for small clusters and
$-1/r$ for larger ones. In the vicinity of the cluster surface, 
the exponentially
decaying $v_{xc}(r)$ will thus be shallower than the image potential, which
then approaches zero much more slowly after a crossover point with the 
LDA potential (see Fig. \ref{im:Na138impot}).

In the spirit of image state calculations for surfaces performed by
Serena {\it et al}. and  Chulkov {\it et at}. \cite{modpot}, 
who apply model potentials 
to correct the erroneous decay of the
Kohn-Sham exchange-correlation potential,
we have constructed an effective potential
\begin{equation}
\label{im:modpot}
  v_{\rm mod}(r,\varepsilon)=\left\{  \begin{array}{lrcl} 
                     \displaystyle v_{xc}(r)
		       & & r & < R_{c}-d \\
		    \displaystyle p(r) & 
		       R_{c}-d < & r & < R_{c}+d \\
                     \displaystyle v_{\rm im}^c(r,\varepsilon)
		       & \quad R_{c}+d < & r & 
                    \end{array} \right. 
\end{equation}
for our jellium clusters, 
based on the classical image potential of a solid sphere (\ref{im:pot}). The
model potential is designed to be local, so that it can be employed 
on the level
of our DFT-LDA calculations. The interpolation function $p(r)$ is a third-order
polynomial which joins smoothly and continuously onto $v_{xc}$ and $v_{\rm im}$
at $r=R_c \pm d$. For the value of $d$ we impose the constraint
$v_{xc}(R_{c}-d)< v_{\rm im}^c(R_{c}+d)$ in order to avoid an unphysical shape of 
the potential in the intermediate region.
Because the occupied wave functions and thus the density might be slightly
altered by the modifications, we determine the model potential
self-consistently, applying Eq.~(\ref{im:modpot}) at every
iteration of the density.

In Fig. \ref{im:Na138impot} we present the model potential obtained for the
Na${}_{138}$ cluster and a dielectric constant of $\varepsilon=1000$. Inside the
cluster, the model potential follows the shape of the LDA-$v_{xc}$ potential, but
in the immediate vicinity of the cluster surface it approaches zero much faster.
At a distance of approximately one-half the cluster radius away from the surface 
the model potential,
which by construction follows the inverse power-law decay of the classical image 
potential, $v_{\rm im}^c(r)$, crosses over with the exponentially decaying
exchange-correlation potential in the LDA, as expected.

The inclusion of long-range correlation effects in this fashion proves to be
necessary to reproduce the same image states as our full
quasiparticle calculations, in particular in those cases where the LDA breaks
down (see Fig. \ref{im:imstates_large}). 
The shape of the image state wave functions and the
energies dependend on the dielectric constant, while the {\it number} of such
states is insensitive to it.

To close the discussion on our model potential, we apply it to a cluster
approaching the mesoscopic size range. For this purpose we chose the 
cluster Na${}_{508}$, that has a radius of 31.9 a.u.. Our model 
potential calculation yields six image
states, whereas only three of them have a corresponding state in the LDA. The
highest image state predicted by the model is very narrowly bound 
($-$0.004 eV). It extends extremely far into the vacuum and reaches its 
intensity maximum as far as two cluster radii away from the surface. This 
observation 
corroborates the conjecture that
the number of image states increases for larger clusters until the familiar
Rydberg series is recovered in the limit of infinitely large clusters. It
further suggests that clusters with 2.5 times this radius, as those studied
experimentally by Kasperovich
{\it et al}. \cite{Kasperovich/Kresin:2000} (see also line for 86.2 a.u. 
in Fig. \ref{im:sphimpot}) will already bind a 
considerable number of image states, which will then
contribute noticeably to the electron capture rate, as observed in the
experiment \cite{Kasperovich/Kresin:2000}.

%
\section{Conclusions}
In conclusion, we have presented image states for small clusters from a full 
quantum-mechanical many-body
calculation. In contrast to surfaces, nanoclusters contain a
finite number of image states, that are subject to quantum confinement effects.
In order to describe image states in the $GW$ approximation correctly, a full
diagonalization of the quasiparticle Hamiltonian is necessary, because the LDA
wave functions no longer provide a good description of the quasiparticle
wave functions.
To extend the discussion to mesoscopic clusters, we have devised a model 
potential that captures the correct asymptotic decay of the image potential
and yields images states in qualitative agreement with the quasiparticle states
of our $GW$ approach. 

\begin{acknowledgments}
We thank Wolf-Dieter Sch\"one, Peter Bokes, Tim Gould, and H\'ector Mera for many fruitful
discussions. This work was supported by the EPSRC and by the EU through the NANOPHASE Research
Training Network (Contract No. HPRN-CT-2000-00167) and in part by the NANOQUANTA Network
of Excellence (NMP4-CT-2004-500198). Patrick Rinke also acknowledges the support of 
the DAAD.
\end{acknowledgments}


\begin{thebibliography}{38}
\bibitem[{foo({\natexlab{a}})}]{footnote:au}
\bibinfo{note}{Atomic units are used throughout this paper.}
\bibitem[{foo({\natexlab{b}})}]{footnote:def_of_imstat}
\bibinfo{note}{An ideal image state is bound by the image potential.
  We therefore
  identify any bound state that is predominantly localized in the vacuum and
  that has an insignifcant overlap with the cluster region as an image state.}
\end{thebibliography}

\end{document}